\documentclass{cernrep}
\usepackage{subfigure}
\usepackage{lineno}
\usepackage{xspace}
\usepackage{hyperref}
\usepackage{graphicx}
\usepackage{lineno}
\usepackage{color}
\begin{document}
\title{Charmonium production as a function of charged-particle multiplicity in pp and p--Pb collisions with ALICE at the LHC}
\author{Theraa TORK, for the ALICE collaboration}
\institute{Laboratoire de Physique des 2 Infinis, Irène Joliot-Curie, Orsay, France\\
Presented at DIS2022: XXIX International Workshop on Deep-Inelastic Scattering and Related Subjects, Santiago de Compostela, Spain, May 2-6 2022.}

\begin{abstract}
Heavy quarkonium production is an excellent tool to test both perturbative and non-perturbative QCD, as perturbative QCD can describe the heavy quark production process, while the formation of the quarkonium bound state involves non-perturbative aspects. In ultrarelativistic heavy-ion collisions, a deconfined state of QCD matter, made of free quarks and gluons, called quark-gluon plasma (QGP) is expected to be formed. To probe such an environment, the study of quarkonium production is an important tool as the heavy quarks are produced during the initial stage of the collision and experience the entire medium evolution. Surprisingly, in small colliding systems as pp and p--Pb, QGP-like behaviours are observed when selecting high multiplicity events. The physics interpretation of these behaviours remains unclear. However, multiparton interaction is one of the main promising scenarios to explain such observation. Studies of charmonium yields as a function of the event charged-particle pseudorapidity density in pp and p--Pb collision allow one to probe multiple parton interactions in an indirect way. These proceedings present the measurements of quarkonium yields normalised to their average values as a function of the charged-particle multiplicity in pp collisions at $\sqrt{s}$ = 13 TeV and in p--Pb collisions at $\sqrt{s_{\rm NN}}$ = 8.16 TeV, performed by the ALICE experiment at the LHC. The corresponding results for the $\psi\rm{(2S)}$-to-J/$\psi$ ratios as a function of charged-particle multiplicity are also shown. In addition, J/$\psi$ pair production measurement in pp collisions at $\sqrt{s}$ = 13 TeV is discussed. 
\end{abstract}


\maketitle

\section{Introduction}
The early stages of a heavy-ion collision (A-A) include the formation of hot and dense matter, known as quark-gluon plasma (QGP), consisting of freely moving quarks and gluons. Quarkonia, i.e. charmonia (c$\Bar{\rm{c}}$) and bottomonia (b$\Bar{\rm{b}}$), are expected to probe the whole evolution of the QGP due to the early production of their constituent quarks. Moreover, quarkonia are an excellent tool to test our current knowledge of QCD as heavy quark production takes place during the hard processes of the collision, whereas the quarkonium bound state formation is a soft process. Consequently, quarkonium production is sensitive to  both perturbative and non-perturbative aspects of the QCD. The LHC data revealed several unexpected behaviours in small systems (pp and p--Pb collisions) at high multiplicity, e.g, the non-zero elliptic flow of identified hadrons through
long-range angular correlation measurementsin p--Pb collisions at $\sqrt{s_{\rm NN}}$ = 5.02 TeV  \cite{ALICE:2013snk} and the enhanced production of multi-strange hadrons in pp collisions at $\sqrt{s}$ = 7 TeV \cite{ALICE:2016fzo}. These findings are
quite surprising, as they are usually interpreted as signatures of the QGP in A--A
collisions. Multiparton interactions (MPI) are one of the main scenarios proposed to explain these observations in small systems. MPIs occur in events in which several parton-parton interactions take place in a single hadron-hadron collision. Several tools can be used to probe MPI with quarkonia, such as double quarkonium production, or quarkonium production as a function of charged-particle multiplicity. The former is a direct probe for MPI, as it is sensitive to double hard parton scatterings. Moreover, it provides information on single quarkonium production mechanisms \cite{Szczurek:2012kt}. However, a large integrated luminosity is required to perform such analysis. On the other hand, the quarkonium production as a function of charged-particle multiplicity is an indirect probe for MPI. It is sensitive to the interplay between soft and hard QCD processes in the event, as quarkonium production is a hard QCD process, whereas the charged particles are usually produced during soft QCD processes. 

\noindent In these proceedings, the results from pp and p--Pb collisions for the self-normalised quarkonium yields, measured at midrapidity ( $|y\rm{_{lab}}|$ < 0.9 ) or forward rapidity ( 2.5 < $|y\rm{_{lab}}|$ < 4.0), as a function of the self-normalised charged-particle multiplicity, measured at midrapidity, are discussed. The ratio of the self-normalised $\psi\rm{(2S)}$-to-J/$\psi$ yields as a function of charged-particle multiplicity, in both pp and p--Pb collisions, are also shown. Finally, the measurement of the J/$\psi$ pair production cross section in pp collisions is presented.

\section{Experimental setup}
ALICE (A Large Ion Collider Experiment) is a general purpose
detector at the LHC, devoted to heavy-ion physics. In addition, it has a rich program in small collision systems, such as pp and p-Pb. ALICE is equipped with 18 different sub-detectors, which can be classified into three main categories, according to their usage in the analyses described below: (i) the muon spectrometer, which reconstructs and identifies the muon tracks at large rapidity, (ii) the central barrel detectors which reconstruct at midrapidity the primary vertex and charged-particle tracks, as well as J/$\psi$ in the dielectron decay channel, (iii) global detectors which contribute to triggering, background rejection and charged-particle multiplicity measurements. More details about the ALICE detector and its performance can be found in Ref.\cite{ALICE:2014sbx}

\section{Results}

\subsection{J/\texorpdfstring{$\psi$}. pair production in pp collisions at \texorpdfstring{$\sqrt{s}$}. = 13 TeV}

\begin{figure}[!htbp]
 \centering
    \subfigure[J/$\psi$ pair production]{
    \includegraphics[width =0.45\textwidth, height= 7cm]{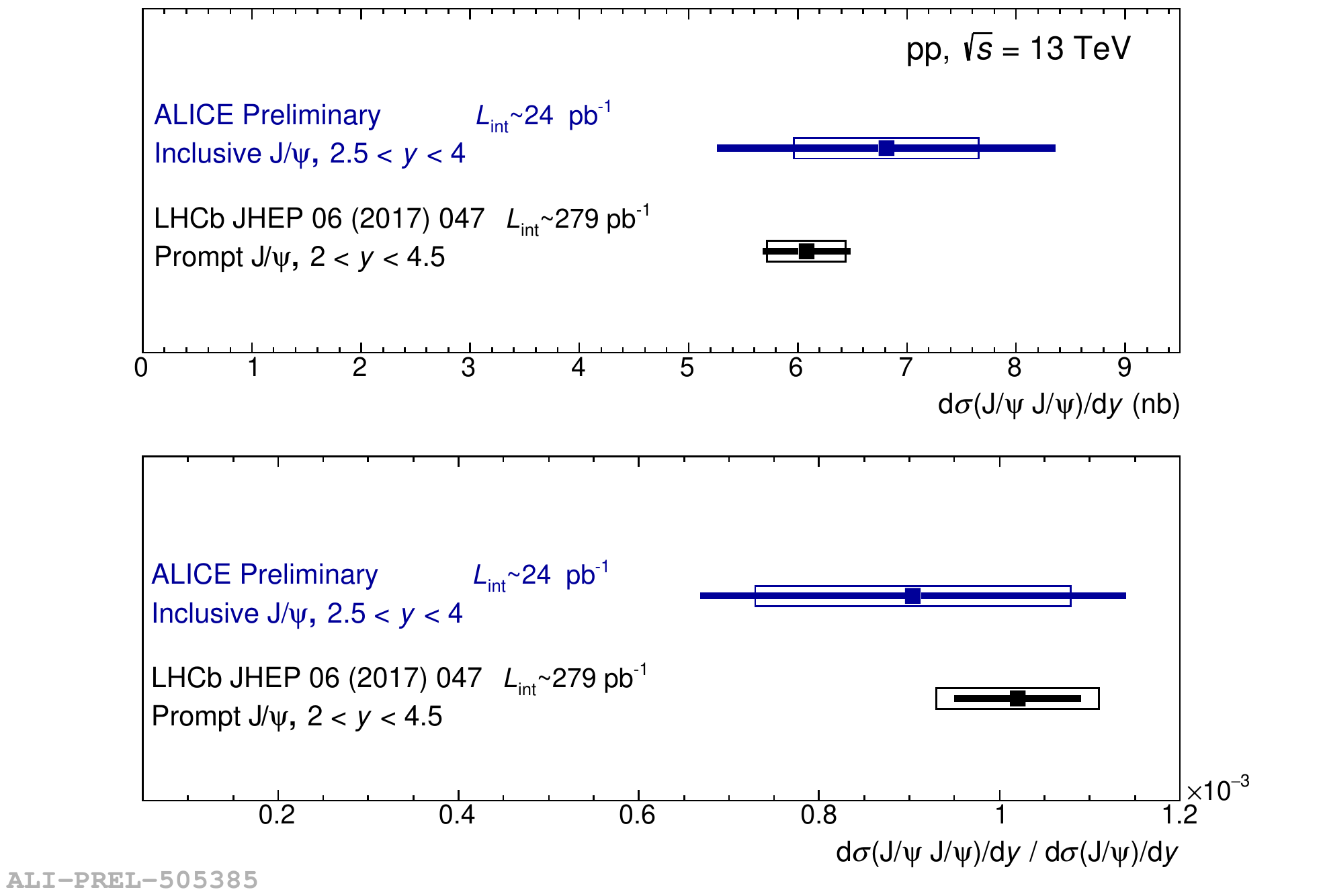}
    \label{fig:doubleJPsi}
  }
    \qquad 
    \subfigure[J/$\psi$ self-normalised yields at midrapidity]{
    \includegraphics[width =0.45\textwidth, height=7cm]{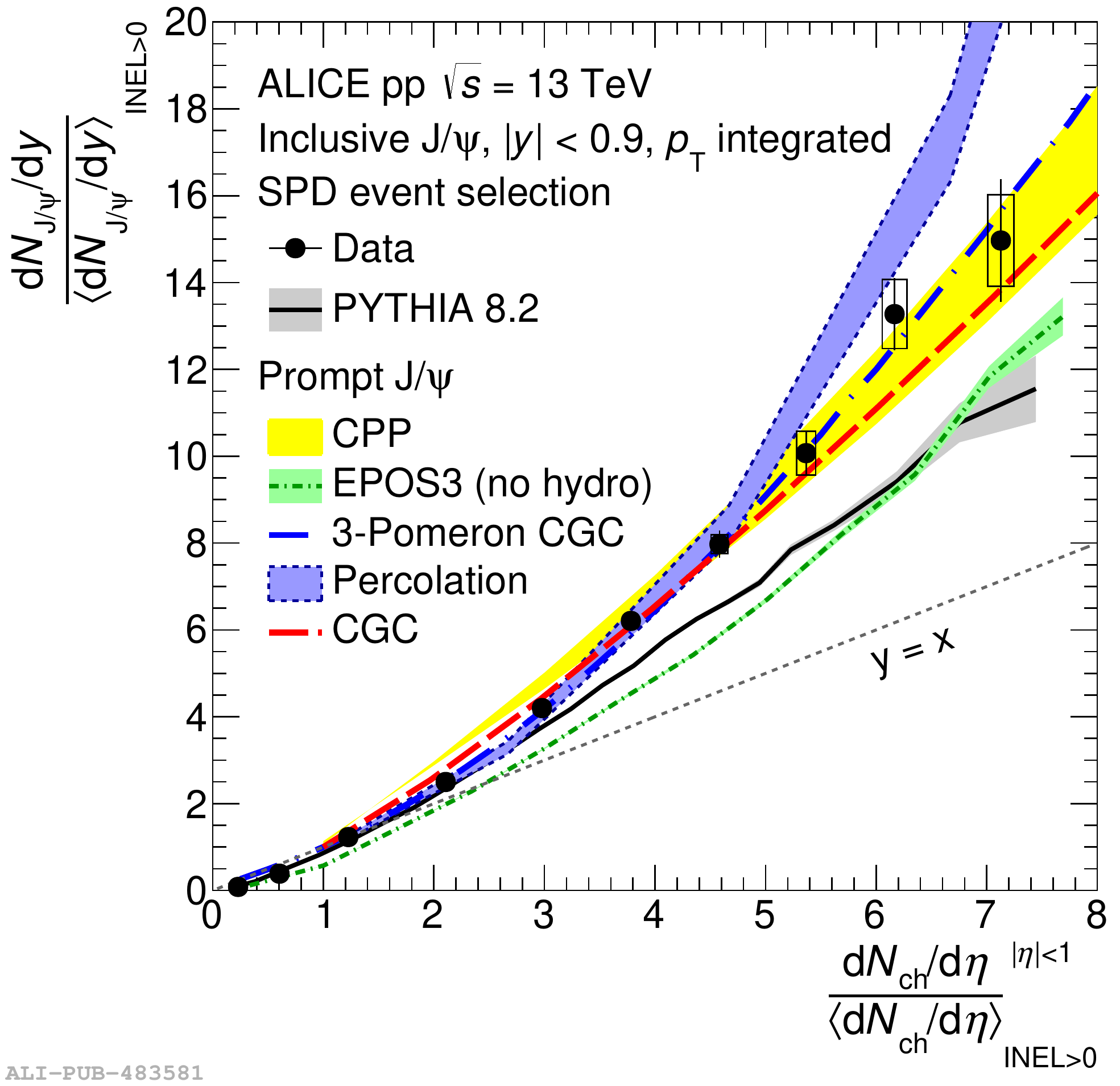}
    \label{fig:JPSi_mid}
  }
    \caption{(a) Inclusive J/$\psi$ pair production cross section (top panel), and ratio of double-to-single inclusive J/$\psi$ production cross sections (bottom panel), in pp collisions at $\sqrt{s}$ = 13 TeV. Results are compared with LHCb measurements for prompt J/$\psi$ production \cite{LHCb:2016wuo}. (b) J/$\psi$ self-normalised yields as a function of the self-normalised charged-particle multiplicity \cite{ALICE:2020msa}. Both quantities are measured at midrapidity. The results are compared with several theoretical models \cite{Kopeliovich:2013yfa,Levin:2019fvb,Ma:2014mri,Ferreiro:2012fb,Sjostrand:2014zea,Drescher:2000ha}.
    }
    \label{fig:pp_odubleandmultiplicity}
\end{figure}

The inclusive J/$\psi$ pair production cross section per rapidity unit in pp collisions at $\sqrt{s}$ = 13 TeV is presented in the top panel of Fig.\ref{fig:doubleJPsi}. The results are compared with those for prompt J/$\psi$ production from the LHCb experiment,  performed in a slightly larger rapidity window\cite{LHCb:2016wuo}. A good agreement is observed between the two measurements within uncertainties. The derived double-to-single J/$\psi$ production cross section results are shown in the bottom panel of the figure. The ALICE and LHCb results are also consistent within the uncertainties.

\subsection{Quarkonium production as a function of charged-particle multiplicity in pp collisions}

The multiplicity dependence of the inclusive J/$\psi$ production, measured at midrapidity, in pp collisions at \mbox{$\sqrt{s}$ = 13 TeV} was reported in Ref.\cite{ALICE:2020msa}.
In this work, the J/$\psi$ yields and the charged-particle multiplicity are  normalised to their respective average values obtained in the integrated multiplicity interval.
As presented in Fig.\ref{fig:JPSi_mid}, the self-normalised J/$\psi$ yields at midrapidity exhibit a faster than a linear increase with the increasing charged-particle multiplicity at midrapidity. The trend of the data is described quantitatively by the following models: the coherent particle production model (CPP) \cite{Kopeliovich:2013yfa}, the color glass condensate model (CGC) \cite{Ma:2014mri},  the 3-pomeron CGC model \cite{Levin:2019fvb}. The percolation model \cite{Ferreiro:2012fb} describes the trend of the data at low multiplicity. The aforementioned models include initial and final state effects, as well as MPI to describe the behavior of the J/$\psi$ yields as a function of the event multiplicity. EPOS3 \cite{Drescher:2000ha} and PYTHIA 8.2 \cite{Sjostrand:2014zea} events generators describe qualitatively a faster than a linear increase but not as strong as in the data.

The J/$\psi$ production at forward rapidity as a function of the charged-particle multiplicity measured at midrapidity in pp collisions at $\sqrt{s}$ = 13 TeV has also been reported \cite{ALICE:2021zkd}.
The J/$\psi$ self-normalised yields show an increase compatible with a linear trend within uncertainties. This increase is therefore less rapid than that of the J/$\psi$ yields measured at midrapidity. 
The origin of the difference observed for these trends as a function of multiplicity remains unclear, although the exercise consisting in varying the rapidity range used to select the charged-particle multiplicity, which is described in Ref.\cite{ALICE:2020msa},  suggests that it is not due to a possible bias when the hard particle is produced in a jet.
 
\begin{figure}[!htbp]
 \centering
    \subfigure[$\psi(2S)$ self-normalised yield.]{
    \includegraphics[width =0.45\textwidth]{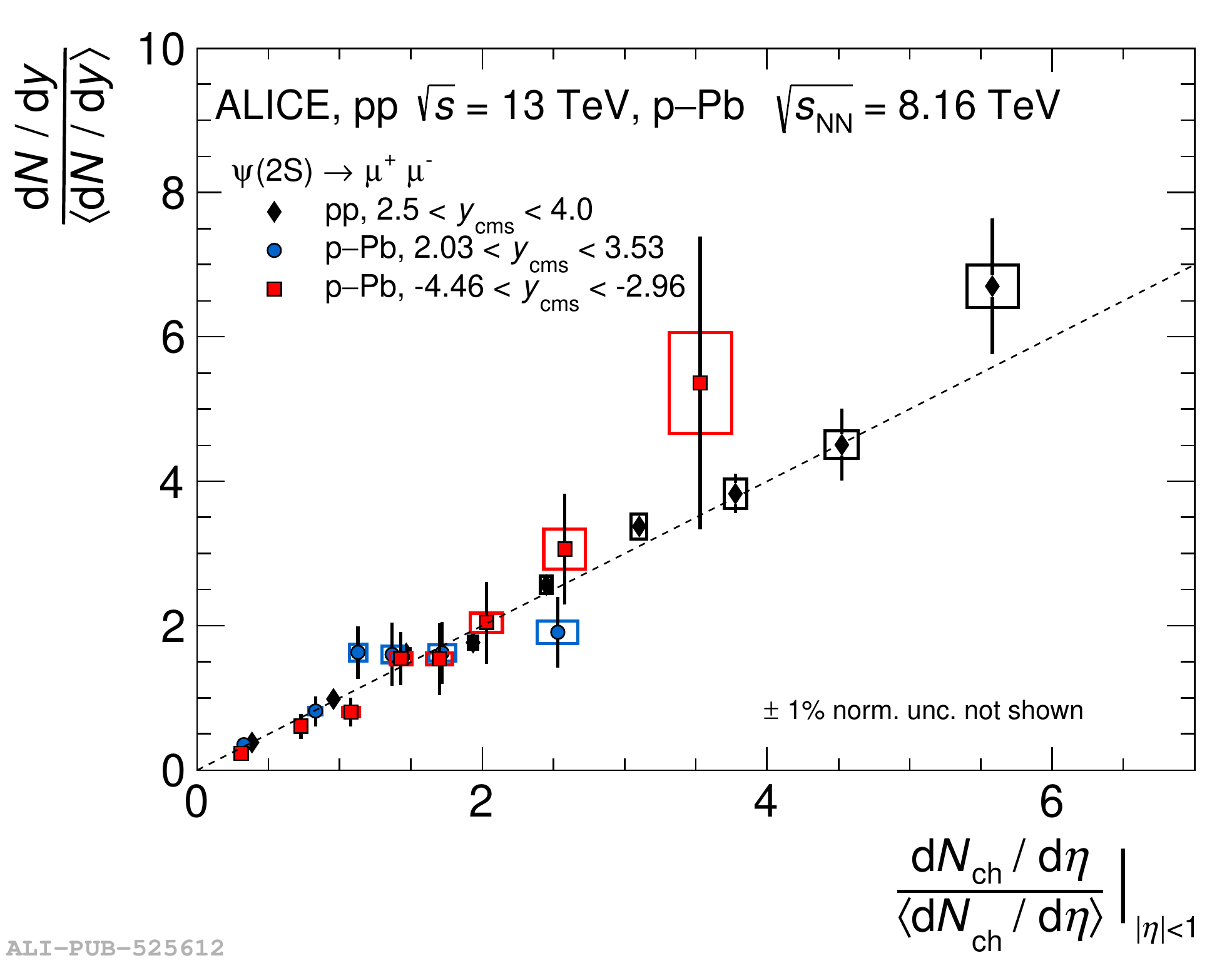}
    \label{fig:Psi2sppAndPpb}
    }
    \qquad
    \subfigure[$\psi(2S)$-to-J/$\psi$ self-normalised yield ratio.]{
    \includegraphics[width =0.45\textwidth]{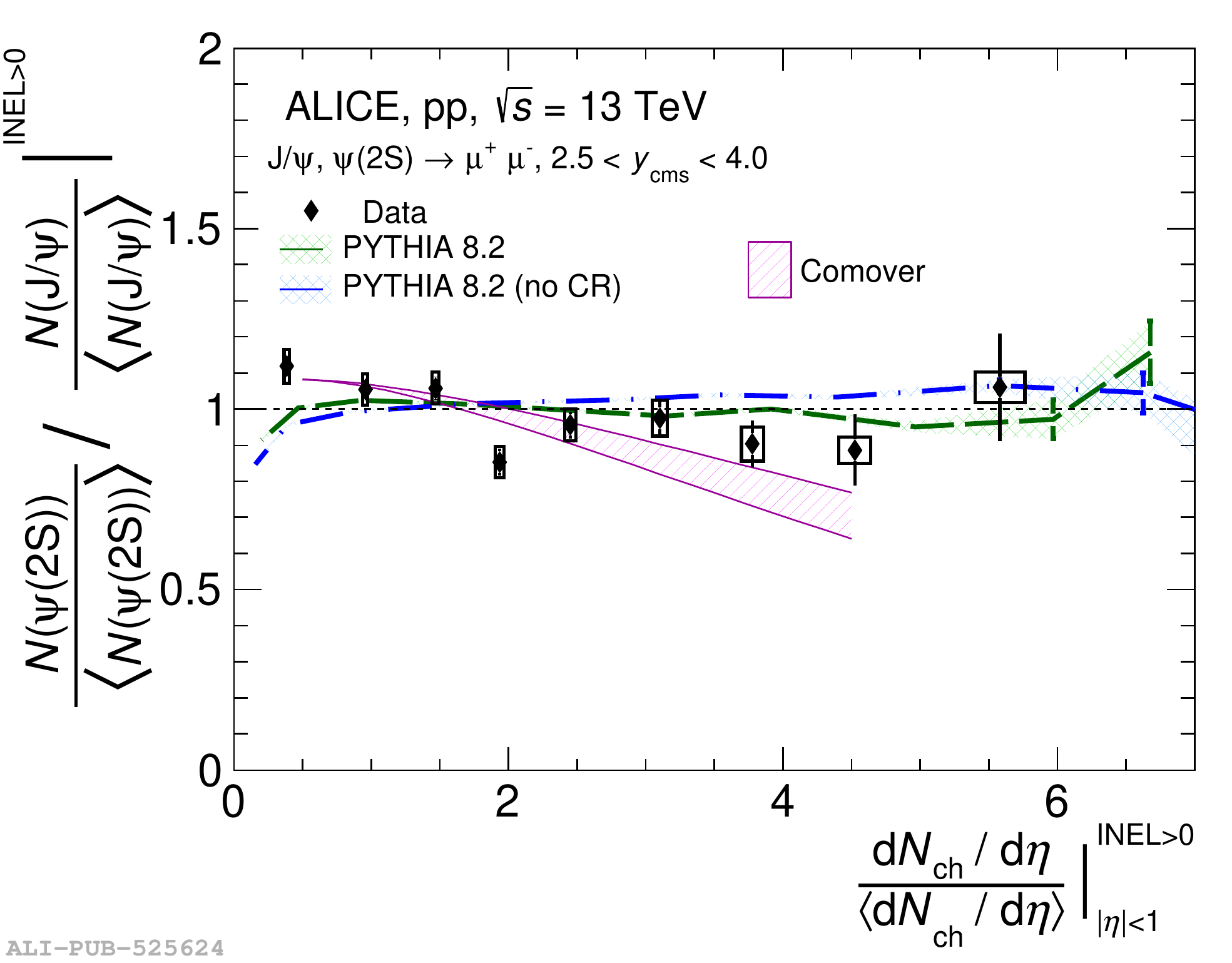}
    \label{fig:psi2sOverJPsiPP}
  }

    \caption{
    (a) Inclusive $\psi(2S)$ self-normalised yields, measured at forward rapidity, as a function of the self-normalised charged-particle multiplicity measured at midrapidity, in pp collisions at $\sqrt{s} = 13$~TeV and in p--Pb (Pb--p) collisions at $\sqrt{s_{\rm NN}} = 8.16$~TeV at forward (backward) rapidity\cite{ALICE:2022gpu}. 
    (b) $\psi(2S)$-to-J/$\psi$ self-normalised yield ratio, measured at forward rapidity, as a function of the self-normalised charged-particle multiplicity, measured at midrapidity, in pp collisions at $\sqrt{s} = 13$~TeV \cite{ALICE:2022gpu}. Results are compared with predictions from comovers model\cite{Ferreiro:2012fb} and PYTHIA 8.2\cite{Sjostrand:2014zea}.
    }
\end{figure}

 \noindent In Ref.\cite{ALICE:2022gpu}, the self-normalised forward $\psi\rm{(2S)}$ yields as a function of the charged-particle multiplicity in pp collisions at $\sqrt{s}$ = 13 TeV are presented. The yields show an approximately linear increase with multiplicity within the uncertainties, see Fig.\ref{fig:Psi2sppAndPpb}. The data are compatible with PYTHIA 8.2 calculations \cite{Sjostrand:2014zea}, both with and without color reconnection scenario (CR), within the uncertainties (see Ref.\cite{ALICE:2022gpu} for the comparison with models). In Fig.\ref{fig:psi2sOverJPsiPP}, the self-normalised $\psi\rm{(2S)}$-to-J/$\psi$ ratio is shown as a function of charged-particle multiplicity. Measuring the excited-to-ground charmonium state yield ratio allows disentangling possible final state effects at play in pp collisions, as most of the other effects are expected to cancel in the ratio. The data show a flat distribution as a function of charged-particle multiplicity. \mbox{PYTHIA 8.2} calculations suggest a similar behavior for both $\psi\rm{(2S)}$ and J/$\psi$. The $\psi\rm{(2S)}$-to-J/$\psi$ self-normalised yields are also compared with the comover model calculations \cite{Ferreiro:2014bia}. The comover model considers the possibility for quarkonia to dissociate due to their interactions with the final state particles that are comoving with them. The dissociation probability depends on the size (binding radius) of the particle and the comover density. The comover dissociation has a larger influence on $\psi\rm{(2S)}$ with respect to J/$\psi$, due to its larger binding radius. In the multiplicity range where the comover model calculations are available, data and model remain consistent within uncertainties.
 
\subsection{Quarkonium production as a function of charged-particle multiplicity in p--Pb collisions}

Several nuclear effects may influence the production of hadrons in p--Pb collisions, as compared to pp collisions. 
The gluon shadowing \cite{Eskola:2009uj} leads to the suppression of quarkonium production in p--Pb collisions with respect to pp collisions which is described by the modification of the parton distribution functions in the nuclei (nPDF) with respect to isolated nucleons. 
The CGC model \cite{Levin:2019fvb} considers that at sufficiently high energies, the gluons interaction probability increases due to their large  density, reducing effectively their total amount (gluon saturation). This scenario can influence both charged-particle and quarkonium production rates. 
The fully coherent energy loss model \cite{Arleo:2021bpv} suggests the emission of gluons, induced by the medium, coherently from partons in the initial state and from the c$\Bar{c}$ pair in the final state. 
On the other hand, the nuclear absorption \cite{Kluberg:2009wc} and the comovers models \cite{Ferreiro:2014bia} influence the production of heavy mesons due to the interaction with final state particles. The nuclear absorption model predicts the breakup of charmonia due to the interaction with the primordial nucleus. However, this effect is neglected at the LHC energies, as the time needed to cross the nucleus by the c$\Bar{\rm{c}}$ pair is much shorter than the charmonium formation time.

\begin{figure}[!htbp]
 \centering
    \subfigure[Self-normalised $\psi\rm{(2S)}$ yield]{
    \includegraphics[width =0.45\textwidth]{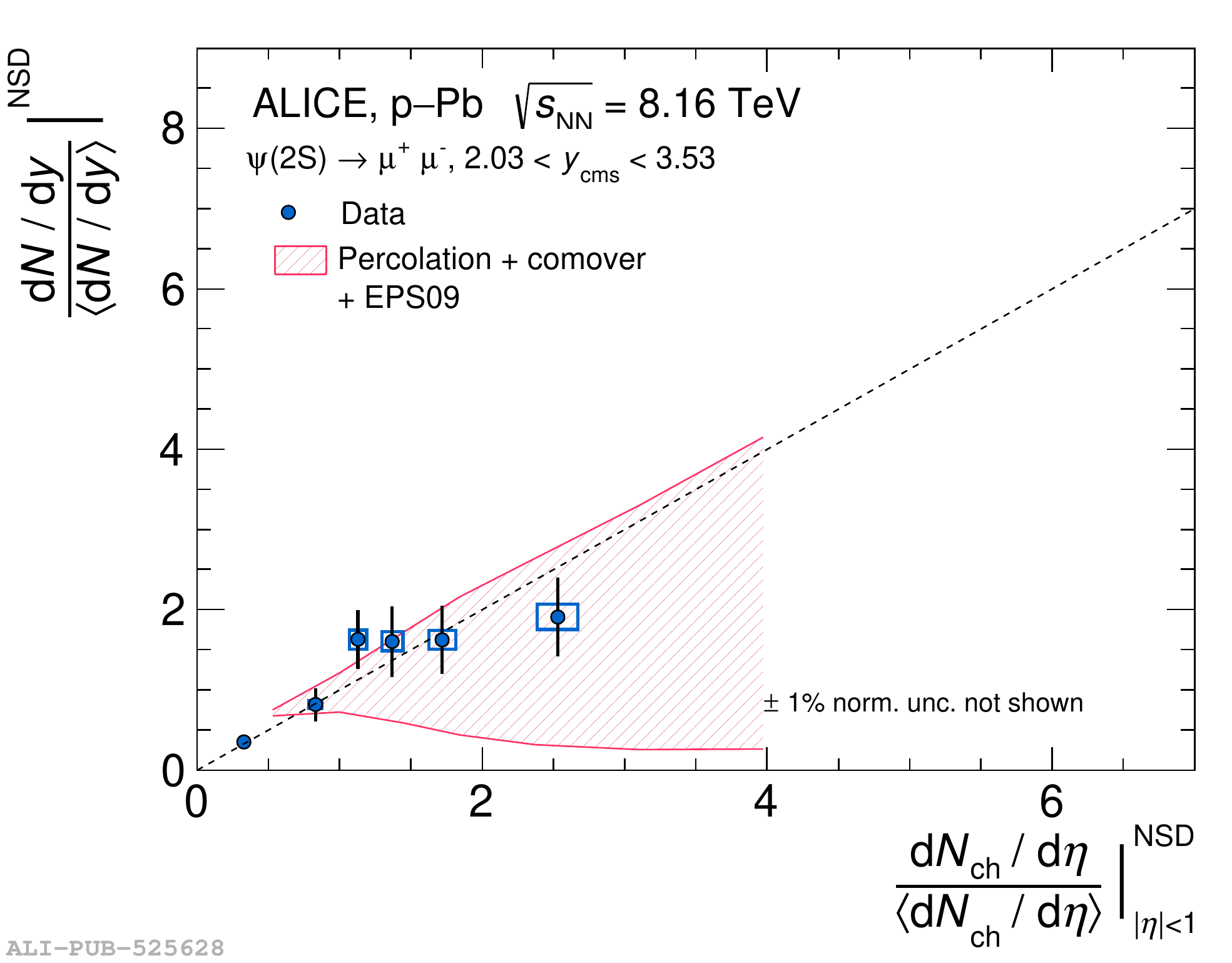}
    \label{fig:psi2SRELFW}
  }
    \qquad 
    \subfigure[Self-normalised $\psi(2S)$-to-J/$\psi$ ratio]{
    \includegraphics[width =0.45\textwidth]{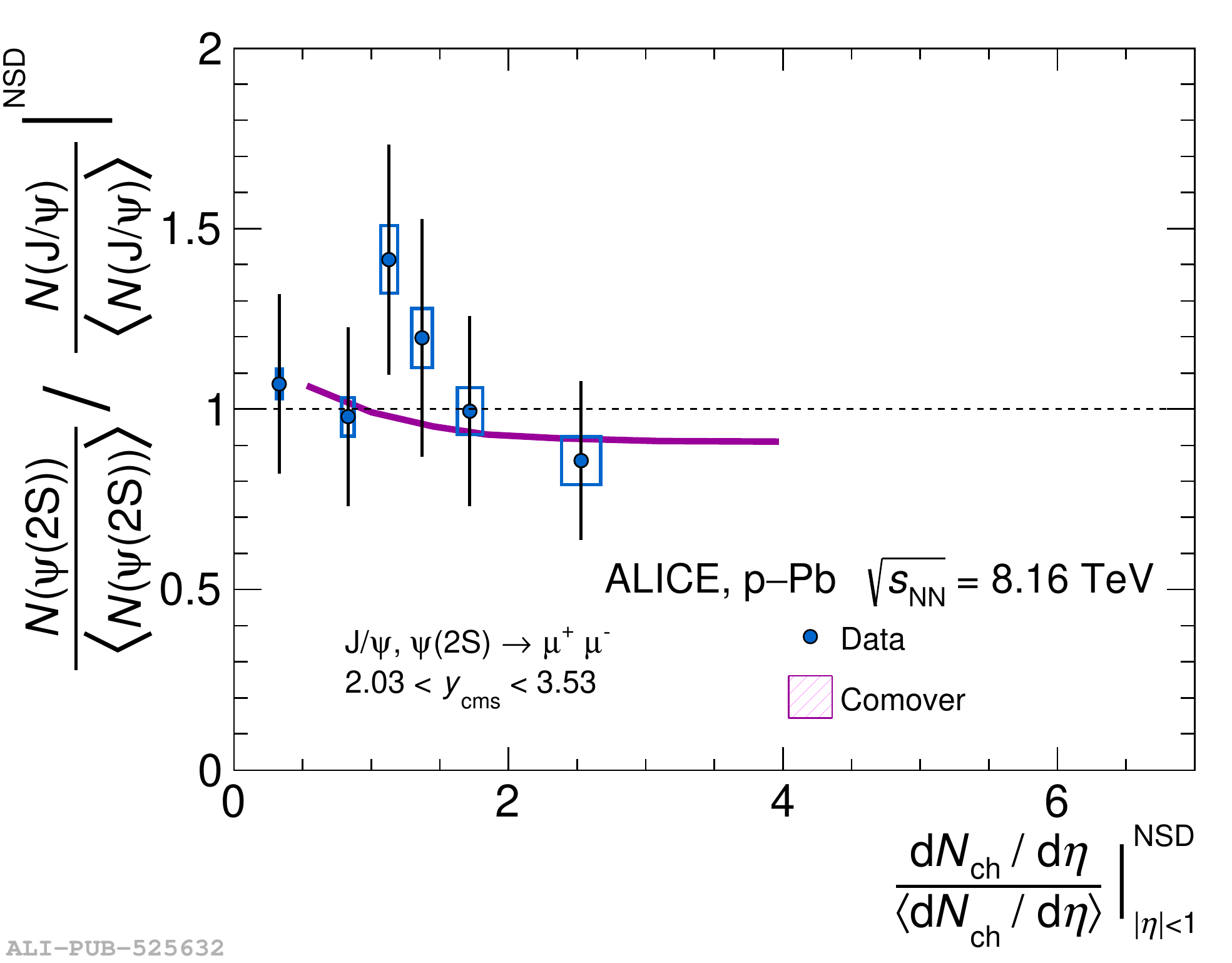}
    \label{fig:ModelPpbfw}
  }
    \caption{(a) The self-normalised $\psi(2S)$ yield and (b) the self-normalised $\psi(2S)$-to-J/$\psi$ yield ratio, measured at forward rapidity, as a function of the charged-particle multiplicity, measured at midrapidity, in p--Pb collisions at $\sqrt{s\rm{_{NN}}}$ = 8.16 TeV \cite{ALICE:2022gpu}. The $\psi(2S)$-to-J/$\psi$ yield ratios are compared to the percolation + comover + EPS09 model within the large uncertainties \cite{Ferreiro:2014bia,Ferreiro:2012fb,Eskola:2009uj}, while the $\psi(2S)$ yields are compared to the comover model \cite{Ferreiro:2014bia}.
    \label{fig:psi2pPb}
    }
\end{figure}

\noindent 
The multiplicity dependence of the J/$\psi$ self-normalised yields in p--Pb collisions at $\sqrt{s\rm{_{NN}}}$ = 8.16 TeV is reported in Ref.\cite{ALICE:2020eji}. In this analysis, J/$\psi$ mesons are reconstructed at forward and backward rapidity in the nucleon-nucleon center-of-mass frame, while the charged-particle multiplicity is measured at midrapidity. The results show an increase of the yields with increasing multiplicity at both forward and backward rapidity. This increase is weaker for the yields reconstructed at forward rapidity with respect to the ones reconstructed at backward rapidity. The trend of the yields are consistent with EPOS3 calculations \cite{Drescher:2000ha}.
The self-normalised yields of the $\psi\rm{(2S)}$ as a function of the self-normalised charged-particle multiplicity in p--Pb collisions are presented in Fig.\ref{fig:Psi2sppAndPpb} and compared to pp collisions. The $\psi\rm{(2S)}$ yield increases as a function of multiplicity at both forward and backward rapidity in p--Pb collisions. This increase is compatible with a linear increase, depicted as a dashed line in the figure, within uncertainties. The $\psi\rm{(2S)}$ self-normalised yields exhibit a similar behaviour in pp and p--Pb collisions within uncertainties. In Fig.\ref{fig:psi2SRELFW}, the self-normalised yields of $\psi\rm{(2S)}$, measured at forward rapidity, are compared with the percolation  coupled to comover model \cite{Ferreiro:2014bia,Ferreiro:2012fb} and using EPS09 nPDF parameterizations \cite{Eskola:2009uj}. 
The model includes the influence of percolation, where the initial state partons are represented as strings with a certain transverse size and can interact with each other. In a dense environment, these strings can overlap, reducing the effective number of strings, and leading to an effective reduction in the quarkonium production. As described before, the comover model is based on the probability for quarkonia to dissociate due to their interactions with final state particles. The percolation + comover + EPS09 model is compatible with the data within uncertainties (the large model uncertainties are dominated by the EPS09 ones). The $\psi(2S)$-to-J/$\psi$ self-normalised yield ratio at forward rapidity as a function of the charged-particle multiplicity at midrapidity is presented in Fig.\ref{fig:ModelPpbfw}  \cite{ALICE:2022gpu}. The result shows a distribution compatible with unity within uncertainties. The measurements are compared to the comovers model calculations \cite{Ferreiro:2014bia}. Within the large experimental uncertainties, the results remain consistent with the comover scenario, which suggests a slightly stronger suppression of the $\psi\rm{(2S)}$ with respect to J/$\psi$ in the high multiplicity region.

\section{Summary}
The  J/$\psi$ pair production in pp collisions at $\sqrt{s}$ = 13 TeV is presented. The result shows a good agreement within uncertainties with similar measurements performed by the LHCb experiment. The J/$\psi$ and $\psi(2S)$ self-normalised yields show an increase as a function of the charged-particle multiplicity in pp collisions at \mbox{$\sqrt{s}$ = 13 TeV} and p--Pb collisions at $\sqrt{s\rm{_{NN}}}$ = 8.16 TeV. The trend of the data is described by several models which include initial and final state effects, as well as MPI. In the measurement of the $\psi(2S)$-to-J/$\psi$ production ratios, a similar behaviour for the two states is observed as a function of multiplicity, within uncertainties. The experimental precision on the excited state measurements together with the model uncertainties do not allow one to disentangle possible final state effects at play. The LHC Run 3 will provide higher luminosity, leading to more precise multiplicity-differential measurements, bringing further constraints to MPI modeling.

{\small
\bibliographystyle{utphys}   
\bibliography{cernrepexa}

\providecommand{\href}[2]{#2}\begingroup\raggedright\begin{thebibliography}{10}

\bibitem{ALICE:2013snk}
{\bfseries ALICE} Collaboration, B.~B. Abelev {\em et~al.}, ``{Long-range
  angular correlations of $\rm \pi$, K and p in p-Pb collisions at
  $\sqrt{s_{\rm NN}}$ = 5.02 TeV}'',
  \href{http://dx.doi.org/10.1016/j.physletb.2013.08.024}{{\em Phys. Lett. B}
  {\bfseries 726} (2013) 164--177},
  \href{http://arxiv.org/abs/1307.3237}{{\ttfamily arXiv:1307.3237 [nucl-ex]}}.

\bibitem{ALICE:2016fzo}
{\bfseries ALICE} Collaboration, J.~Adam {\em et~al.}, ``{Enhanced production
  of multi-strange hadrons in high-multiplicity proton-proton collisions}'',
  \href{http://dx.doi.org/10.1038/nphys4111}{{\em Nature Phys.} {\bfseries 13}
  (2017) 535--539}, \href{http://arxiv.org/abs/1606.07424}{{\ttfamily
  arXiv:1606.07424 [nucl-ex]}}.

\bibitem{Szczurek:2012kt}
A.~Szczurek and R.~Maciula, ``{Production of one and two $c \bar c$ pairs at
  LHC}'', \href{http://dx.doi.org/10.1063/1.4802161}{{\em AIP Conf. Proc.}
  {\bfseries 1523} no.~1, (2013) 255--259},
  \href{http://arxiv.org/abs/1212.5427}{{\ttfamily arXiv:1212.5427 [hep-ph]}}.

\bibitem{ALICE:2014sbx}
{\bfseries ALICE} Collaboration, B.~B. Abelev {\em et~al.}, ``{Performance of
  the ALICE Experiment at the CERN LHC}'',
  \href{http://dx.doi.org/10.1142/S0217751X14300440}{{\em Int. J. Mod. Phys. A}
  {\bfseries 29} (2014) 1430044},
  \href{http://arxiv.org/abs/1402.4476}{{\ttfamily arXiv:1402.4476 [nucl-ex]}}.

\bibitem{LHCb:2016wuo}
{\bfseries LHCb} Collaboration, R.~Aaij {\em et~al.}, ``{Measurement of the
  J/$\psi$ pair production cross-section in pp collisions at $ \sqrt{s}=13 $
  TeV}'', \href{http://dx.doi.org/10.1007/JHEP06(2017)047}{{\em JHEP}
  {\bfseries 06} (2017) 047}, \href{http://arxiv.org/abs/1612.07451}{{\ttfamily
  arXiv:1612.07451 [hep-ex]}}. [Erratum: JHEP 10, 068 (2017)].

\bibitem{ALICE:2020msa}
{\bfseries ALICE} Collaboration, S.~Acharya {\em et~al.}, ``{Multiplicity
  dependence of J/$\psi$ production at midrapidity in pp collisions at
  $\sqrt{s}$ = 13 TeV}'',
  \href{http://dx.doi.org/10.1016/j.physletb.2020.135758}{{\em Phys. Lett. B}
  {\bfseries 810} (2020) 135758},
  \href{http://arxiv.org/abs/2005.11123}{{\ttfamily arXiv:2005.11123
  [nucl-ex]}}.

\bibitem{Kopeliovich:2013yfa}
B.~Z. Kopeliovich, H.~J. Pirner, I.~K. Potashnikova, K.~Reygers, and
  I.~Schmidt, ``{J/\ensuremath{\psi} in high-multiplicity pp collisions:
  Lessons from pA collisions}'',
  \href{http://dx.doi.org/10.1103/PhysRevD.88.116002}{{\em Phys. Rev. D}
  {\bfseries 88} no.~11, (2013) 116002},
  \href{http://arxiv.org/abs/1308.3638}{{\ttfamily arXiv:1308.3638 [hep-ph]}}.

\bibitem{Levin:2019fvb}
E.~Levin, I.~Schmidt, and M.~Siddikov, ``{Multiplicity dependence of quarkonia
  production in the CGC approach}'',
  \href{http://dx.doi.org/10.1140/epjc/s10052-020-8086-4}{{\em Eur. Phys. J. C}
  {\bfseries 80} no.~6, (2020) 560},
  \href{http://arxiv.org/abs/1910.13579}{{\ttfamily arXiv:1910.13579
  [hep-ph]}}.

\bibitem{Ma:2014mri}
Y.-Q. Ma and R.~Venugopalan, ``{Comprehensive Description of
  J/\ensuremath{\psi} Production in Proton-Proton Collisions at Collider
  Energies}'', \href{http://dx.doi.org/10.1103/PhysRevLett.113.192301}{{\em
  Phys. Rev. Lett.} {\bfseries 113} no.~19, (2014) 192301},
  \href{http://arxiv.org/abs/1408.4075}{{\ttfamily arXiv:1408.4075 [hep-ph]}}.

\bibitem{Ferreiro:2012fb}
E.~G. Ferreiro and C.~Pajares, ``{High multiplicity $pp$ events and $J/\psi$
  production at LHC}'',
  \href{http://dx.doi.org/10.1103/PhysRevC.86.034903}{{\em Phys. Rev. C}
  {\bfseries 86} (2012) 034903},
  \href{http://arxiv.org/abs/1203.5936}{{\ttfamily arXiv:1203.5936 [hep-ph]}}.

\bibitem{Sjostrand:2014zea}
T.~Sj\"ostrand, S.~Ask, J.~R. Christiansen, R.~Corke, N.~Desai, P.~Ilten,
  S.~Mrenna, S.~Prestel, C.~O. Rasmussen, and P.~Z. Skands, ``{An introduction
  to PYTHIA 8.2}'', \href{http://dx.doi.org/10.1016/j.cpc.2015.01.024}{{\em
  Comput. Phys. Commun.} {\bfseries 191} (2015) 159--177},
  \href{http://arxiv.org/abs/1410.3012}{{\ttfamily arXiv:1410.3012 [hep-ph]}}.

\bibitem{Drescher:2000ha}
H.~J. Drescher, M.~Hladik, S.~Ostapchenko, T.~Pierog, and K.~Werner, ``{Parton
  based Gribov-Regge theory}'',
  \href{http://dx.doi.org/10.1016/S0370-1573(00)00122-8}{{\em Phys. Rept.}
  {\bfseries 350} (2001) 93--289},
  \href{http://arxiv.org/abs/hep-ph/0007198}{{\ttfamily arXiv:hep-ph/0007198}}.

\bibitem{ALICE:2021zkd}
{\bfseries ALICE} Collaboration, S.~Acharya {\em et~al.}, ``{Forward rapidity
  J/\ensuremath{\psi} production as a function of charged-particle multiplicity
  in pp collisions at $ \sqrt{s} $ = 5.02 and 13 TeV}'',
  \href{http://dx.doi.org/10.1007/JHEP06(2022)015}{{\em JHEP} {\bfseries 06}
  (2022) 015}, \href{http://arxiv.org/abs/2112.09433}{{\ttfamily
  arXiv:2112.09433 [nucl-ex]}}.

\bibitem{ALICE:2022gpu}
{\bfseries ALICE} Collaboration, ``{Measurement of $\psi$(2S) production as a
  function of charged-particle pseudorapidity density in pp collisions at
  $\sqrt{s}$ = 13 TeV and p-Pb collisions at $\sqrt{s_{\rm NN}}$ = 8.16 TeV
  with ALICE at the LHC}'', \href{http://arxiv.org/abs/2204.10253}{{\ttfamily
  arXiv:2204.10253 [nucl-ex]}}.

\bibitem{Ferreiro:2014bia}
E.~G. Ferreiro, ``{Excited charmonium suppression in proton\textendash{}nucleus
  collisions as a consequence of comovers}'',
  \href{http://dx.doi.org/10.1016/j.physletb.2015.07.066}{{\em Phys. Lett. B}
  {\bfseries 749} (2015) 98--103},
  \href{http://arxiv.org/abs/1411.0549}{{\ttfamily arXiv:1411.0549 [hep-ph]}}.

\bibitem{Eskola:2009uj}
K.~J. Eskola, H.~Paukkunen, and C.~A. Salgado, ``{EPS09: A New Generation of
  NLO and LO Nuclear Parton Distribution Functions}'',
  \href{http://dx.doi.org/10.1088/1126-6708/2009/04/065}{{\em JHEP} {\bfseries
  04} (2009) 065}, \href{http://arxiv.org/abs/0902.4154}{{\ttfamily
  arXiv:0902.4154 [hep-ph]}}.

\bibitem{Arleo:2021bpv}
F.~Arleo, G.~Jackson, and S.~Peign\'e, ``{Impact of fully coherent energy loss
  on heavy meson production in pA collisions}'',
  \href{http://dx.doi.org/10.1007/JHEP01(2022)164}{{\em JHEP} {\bfseries 01}
  (2022) 164}, \href{http://arxiv.org/abs/2107.05871}{{\ttfamily
  arXiv:2107.05871 [hep-ph]}}.

\bibitem{Kluberg:2009wc}
L.~Kluberg and H.~Satz, {\em {Color Deconfinement and Charmonium Production in
  Nuclear Collisions}}.
\newblock 2010.
\newblock \href{http://arxiv.org/abs/0901.3831}{{\ttfamily arXiv:0901.3831
  [hep-ph]}}.

\bibitem{ALICE:2020eji}
{\bfseries ALICE} Collaboration, S.~Acharya {\em et~al.}, ``{J/$\psi$
  production as a function of charged-particle multiplicity in p-Pb collisions
  at $\sqrt{\textit{s}_{\rm NN}}~=~8.16$ TeV}'',
  \href{http://dx.doi.org/10.1007/JHEP09(2020)162}{{\em JHEP} {\bfseries 09}
  (2020) 162}, \href{http://arxiv.org/abs/2004.12673}{{\ttfamily
  arXiv:2004.12673 [nucl-ex]}}.

\end{thebibliography}\endgroup
}

\end{document}